# Silver Amalgam Nanoparticles and Microparticles: A Novel Plasmonic Platform for Spectroelectrochemistry

Filip Ligmajer,*,†,‡,§ Michal Horák,†,§ Tomáš Šikola,†,§ Miroslav Fojta,‡,∥ Aleš Daňhel‡

† *Central European Institute of Technology, Brno University of Technology, Purkyňova 123, Brno, 612 00, Czech Republic*

‡ *Institute of Biophysics of the Czech Academy of Sciences, Královopolská 135, Brno, 612 65, Czech Republic*

§ *Institute of Physical Engineering, Faculty of Mechanical Engineering, Brno University of Technology, Technická 2, Brno, 616 69, Czech Republic*

∥ *Central European Institute of Technology, Masaryk University, Kamenice 753/5, Brno, 625 00, Czech Republic*

**ABSTRACT**

Plasmonic nanoparticles from unconventional materials can improve or even bring some novel functionalities into the disciplines inherently related to plasmonics such as photochemistry or (spectro)electrochemistry. They can, for example, catalyze various chemical reactions or act as nanoelectrodes and optical transducers in various applications. Silver amalgam is the perfect example of such an unconventional plasmonic material, albeit it is well-known in the field of electrochemistry for its wide cathodic potential window and strong adsorption affinity of biomolecules to its surface. In this study, we investigate in detail the optical properties of nanoparticles and microparticles made from silver amalgam and correlate their plasmonic resonances with their morphology. We use optical spectroscopy techniques on the ensemble level and electron energy loss spectroscopy on the single-particle level to demonstrate the extremely wide spectral range covered by the silver amalgam localized plasmonic resonances, ranging from ultraviolet all the way to the mid-infrared wavelengths. Our results establish silver amalgam as a suitable material for introduction of plasmonic functionalities into photochemical and spectroelectrochemical systems, where the plasmonic enhancement of electromagnetic fields and light emission processes could synergistically meet with the superior electrochemical characteristics of mercury.

**INTRODUCTION**

Silver amalgam (AgA) is one of the most suitable solid electrode materials applicable as an alternative to liquid mercury in electroanalysis of various reducible organic and inorganic compounds. These include heavy metals, agrochemicals, colorants, drugs, environmental pollutants,[1–5] or biologically important compounds such as DNA, proteins and their constituents,[6–8] and vitamins.[9] The main advantage of AgA within this context is its wide cathodic potential window, high mechanical stability, adequate sensitivity, and advantageous strong interaction with biopolymers (e.g., DNA and proteins). Intrinsic redox signals of these biopolymers then allow direct analysis of their structure and can thus be utilized for diagnostic purposes in novel biochemical or biomedical applications.[10,11]

In the recent years, many sorts of electrode architectures have benefited from the emergence of nanotechnology and nanomaterials.[12,13] By the same token, nanostructuring of AgA electrodes holds promise for achieving higher charge transfer efficiencies, larger active surface areas, and even some novel catalytic effects.[14,15] Synthesis of stable and homogeneous AgA nanoparticles (NPs) and microparticles (MPs) with controllable and uniform size is, however, somewhat difficult. The most popular approach resides in direct amalgamation of gold[16–19] or silver[20,21] colloidal NPs, which leads to production of either core–shell or amorphous bimetallic NPs. Both types of these amalgamated NPs have found applications primarily in spectroscopic detection of mercury ions[22–25] and have not been investigated with respect to their electrochemical properties so far. Recently, we have demonstrated a novel method for production of AgA particles on top of conductive supports via controlled electrodeposition from a mixed solution of silver and mercury ions.[26] This co-deposition process results in true AgA particles to which DNA and proteins exhibit a preferential affinity, compared to the underlying indium tin oxide (ITO) or pyrolytic graphite, as demonstrated by experiments involving adsorption of fluorescently labeled DNA and green fluorescent protein, respectively.[27] Unlike electrodes made of amalgamated gold, these true AgA electrodes provide also wider potential window and better charge transfer rates in the electrochemical experiments.[26] Moreover, the nanostructured morphology of these electrodes and their electrocatalytic activity towards reduction of organic nitro compounds[14] render them promising for future spectroelectrochemical applications.

Spectroelectrochemistry is a discipline where optical spectroscopy and electrochemical methods are used together to investigate the nature of chemical processes at the nanoscale[28–33] and also to quantify them for analytical and sensing purposes.[34–36] The AgA NPs — with their inherent metallic nature and unique electrochemical qualities — could be an ideal building block for such an integrated platform. There is, however, a significant lack of knowledge about their optical properties, i.e., their dielectric function, or nature and position of their plasmonic resonances. Here, we present a study devoted to investigation of these features. In the first step, we verify that our AgA NPs and MPs indeed exhibit plasmonic resonances of sufficient quality for spectroelectrochemistry, not only in the visible but also in the ultraviolet and infrared wavelength range. We support our findings with a set of numerical simulations, employing the dielectric function measured on a thick AgA film as a proxy for the material properties of our particles. Finally, we also investigate plasmonic properties at the single-particle level, utilizing electron energy loss spectroscopy (EELS), to elucidate that the AgA NPs can support multiple modes of both spectrally separated and overlapping plasmonic resonances. Our results establish AgA NPs and MPs as promising candidates for applications within photochemistry and spectroelectrochemistry[37–39], where the synergy between their plasmonic and electrochemical qualities can be fully utilized.

**METHODS**

**Preparation of AgAP on ITO**. Silver amalgam nano- and microparticles were prepared using the method described in Ref. 26. Briefly, they were electrodeposited onto indium tin oxide (ITO) substrate electrodes (CG-51IN, Delta Technologies, USA) from a mixture of 0.01 mol.l$^{-1}$ AgNO$_3$ and 0.01 mol.l$^{-1}$ Hg(NO$_3$)$_2$ (54% Ag$^+$ (n/n), as confirmed by inductively coupled plasma mass spectrometry) in 0.1 mol.l$^{-1}$ KNO$_3$. Double pulse chronoamperometry was used for the electrodeposition in a two-electrode setup with silver wire pseudoreference electrode in a Teflon cell. The nucleation pulse potential ($E_1 = -1.0$ V) was applied for the nucleation time ($t_1 = 50$ ms), followed by the growth potential ($E_2 = -0.1$ V) that lasted for the variable growth time ($t_2 = 1$–30 min).

**Preparation of bulk AgA**. The bulk silver solid amalgam was prepared by direct mixing of the mercury (99.999 %, Polarografie Praha, Czech Republic) and silver powder (2–3.5 μm, 99.9 %, Sigma-Aldrich, Germany) in a proper weight ratio (60:40, wHg/wAg) by shaking on a Vortex mixer for 1 min in 2 ml Eppendorf tube. Solid amalgam prepared this way was left to harden overnight, then poured into an EpoFix resin (Struers, USA), and consequently polished by emery papers of grain sizes 1200, 2400, and 4000. Final fine polishing was carried out by a filter paper wetted by deionized water.

**Electron microscopy**. The morphology of the samples was also investigated using a scanning electron microscope (SEM, FEI Verios 460L) equipped with an energy-dispersive X-ray spectroscopy (EDS) system (EDAX SDD Octane Super). The EDS quantification was performed using the peak-to-background ZAF method, which is optimal for EDS on samples with spherical geometry.[40] For transmission electron microscope (TEM) analysis, the AgA NPs were washed off the ITO substrate into demineralized water with the aid of sonication. A few microliters of this suspension were then drop-casted onto a silicon nitride membrane (30 nm thick; Agar Scientific). TEM measurements were performed with an FEI Titan microscope equipped with a GIF Quantum spectrometer for EELS and Super-X spectrometer for EDS operated in a monochromated scanning regime at 300 kV. The beam current was set to 0.4 nA and the full width at half-maximum (FWHM) of the zero-loss peak (ZLP) was around 0.13 eV. In the case of EELS, we set the convergence angle to 10 mrad, collection angle to 6.6 mrad, and dispersion of the spectrometer to 0.01 eV/pixel. We recorded EELS spectral images with the pixel size of 3 nm. Every pixel consists of 10 cross-correlated EEL spectra with the acquisition time of 0.4 ms/spectrum. Presented EEL spectra were integrated over several pixels in the region of interest, ZLP and background were subtracted from them, and they were divided by the integral intensity of the ZLP (energy window ±1 eV) to transform measured counts to a quantity proportional to the loss probability. Presented EEL maps show the integrated intensity at the plasmon peak energy with the energy window of 0.1 eV. In the case of EDS, we integrated approximately 50 spectral maps with the acquisition time of 10 μs per pixel. The EDS quantification in weight percent was performed in Velox software using the parabolic background model and Brown–Powell ionization cross-section model for L edges of Ag and Hg.

**Optical spectroscopy**. The far-field optical reflectance spectra were recorded using an infrared microscope (Bruker Hyperion 3000, 15× objective), which was coupled to a Fourier-transform infrared spectrometer (Bruker Vertex v80). The reflected signal was referenced to a gold mirror. The dielectric function of the bulk AgA was measured at three angles of incidence using a rotating analyzer ellipsometer (J. A. Woollam VASE).

**Theoretical calculations**. The dielectric function of the bulk AgA measured by spectroscopic ellipsometry was used as the material parameter in all calculations. The scattering and extinction coefficients were computed using Lorenz–Mie theory[41] and also verified using commercial finite-difference time-domain (FDTD) solver Lumerical FDTD Solutions. Numerical simulations of EELS spectra were performed using the MNPBEM toolbox,[42] which is based on the boundary element method (BEM).[43] The dielectric function of the silicon nitride membrane was approximated by a constant value of 4 (Ref. [44]). For the calculations of EEL spectra and surface charge distribution, the simulated 300 keV electron beam with a FWHM of 0.1 eV was positioned 10 nm from the outer side of the NP and only surface losses were considered. The obtained loss probability density was recalculated in the form of loss probability within 0.01 eV energy intervals (corresponding to the actual dispersion of the spectrometer used in the experiment).

**RESULTS AND DISCUSSION**

The morphology and composition of our AgA particles are analyzed in Figure 1, where the SEM image and the corresponding EDS maps of both silver and mercury confirm the homogeneous distribution of both elements. The EDS linescan (Figure 1b) reveals the comparably higher content of mercury, which is also verified by a full EDS spectrum (Figure 1c), from which we estimate the mercury weight percentage to (62 ± 4) %.

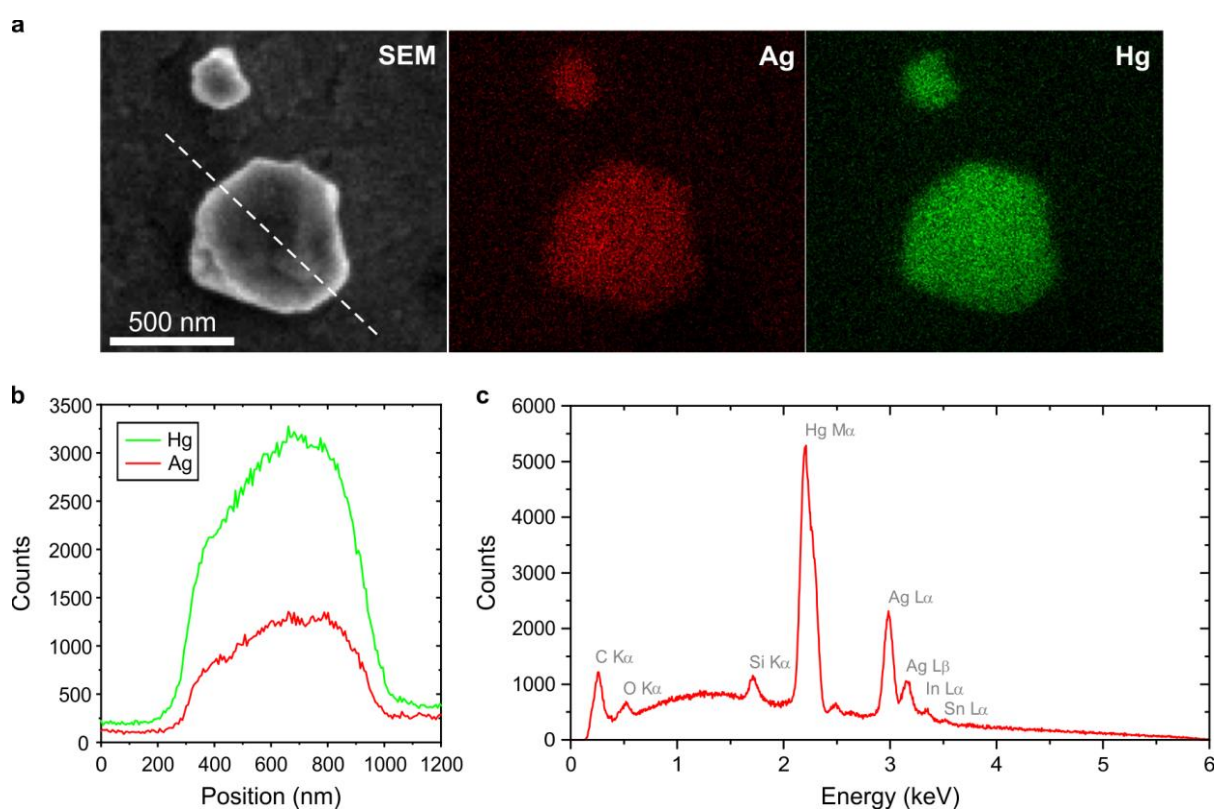

**Figure 1:** (a) SEM image and corresponding Ag and Hg EDS maps of representative AgA particles on top of an ITO substrate. (b) Quantitative Ag and Hg EDS line profiles for the same MP, taken along the white dashed line in the SEM image. (c) Typical full EDS spectrum taken from the center of the same AgA MP.

While the Ag/Hg ratio of the AgA particles can be influenced by the composition of the growth solution,[26] their size, on the other hand, can be controlled by the growth time $t_2$ (Figure 2a–e). For short growth times ($t_2 \leq 10$ min), the size distribution is unimodal, centered on the smallest NPs, with a progressively wider tail towards larger sizes as the growth time increases (Supporting Information, Figure S1). At longer growth times ($t_2 > 10$ min), the size distribution turns into bimodal one, with a large amount of very small AgA NPs accompanied by a small number of very large AgA MPs. Such heterogeneity is often attributed to a process of progressive nucleation during electrodeposition, where the individual particles start to grow at different times.[45] With a wide range of AgA particle sizes at our disposal, we proceeded to the analysis of their optical properties, as such variability should naturally translate into the wide range of plasmonic resonances.[46] The reflectance spectra from the individual AgA electrodes are shown in Figure 2f. Each of them is dominated by a broad peak at mid-infrared (MIR) wavelengths, the distinctive evidence of MIR plasmonic resonances.[47,48] With the increasing average AgA particle size, there is also a characteristic gradual red shift of the resonance frequency, in accordance with the theoretical expectations.[49] Note that the resonances with the highest quality factors can be observed on the samples with intermediate growth times (samples b–d). When the AgA particles grow for too long (sample e), their resulting bimodal size distribution translates into a wide multimodal plasmonic resonance. Although sharp resonances are often necessary for applications in refractive index sensing[50,51] or light modulation,[52] broadband resonances like these here are indispensable especially in plasmon-enhanced MIR vibrational sensing.[53–55] We ascribe the enhanced reflectance peak at low wavenumbers to the large particle clusters visible in Fig. 2e, which naturally have a very high extinction efficiency and thus dominate the reflectance spectra. We would like to note that the variability of the Ag/Hg ratio in particles of different sizes could in principle influence the position of the plasmonic resonances in parallel with the size effect. Although our EDS analysis of the AgA MPs on similar samples indeed revealed some differences in their composition (Supporting Information, Figure S2), the composition differences are below 10% of the Ag/Hg ratio and thus of only limited influence.

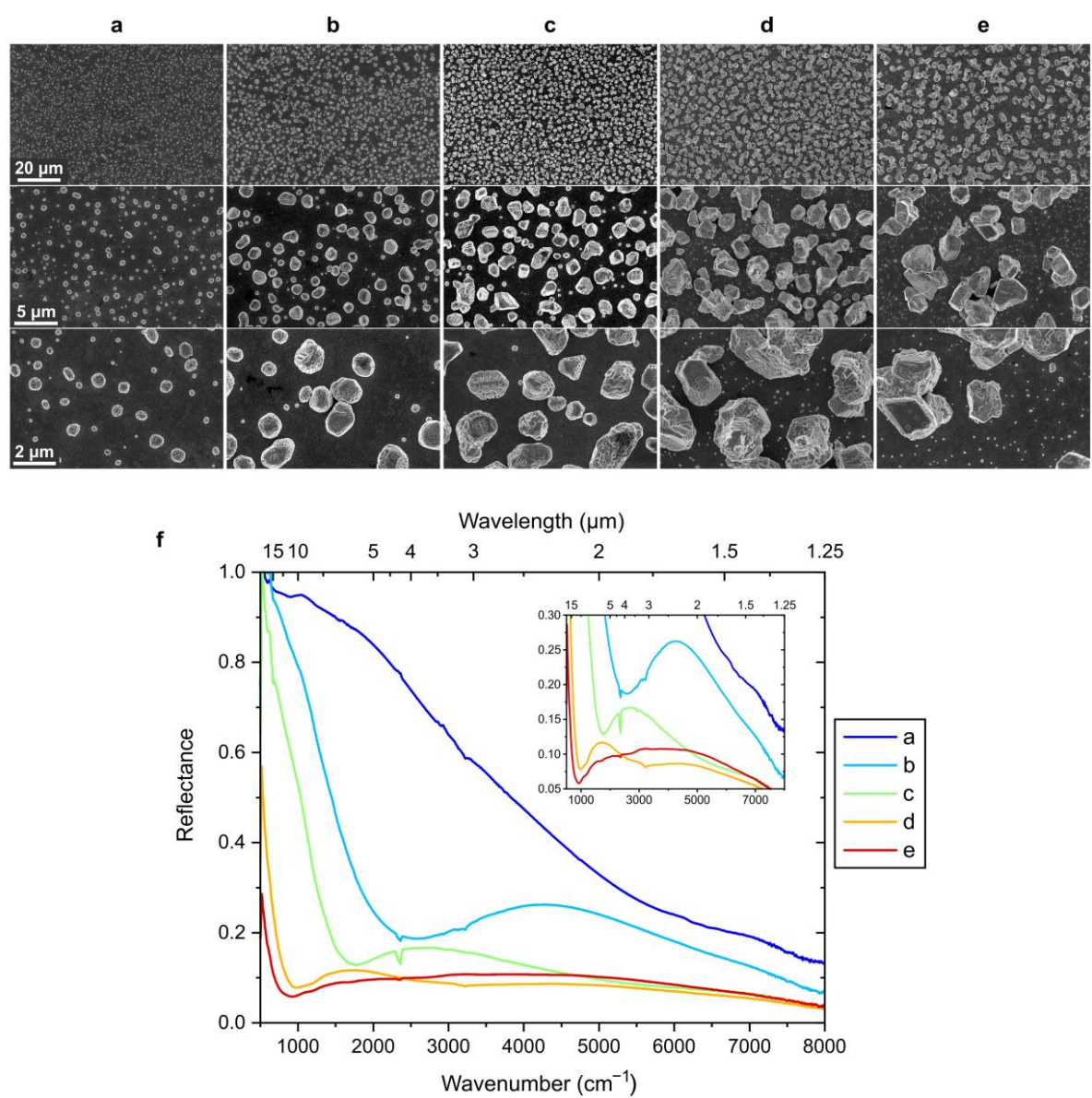

**Figure 2:** Effect of the growth time on the morphology and plasmonic properties of the AgA MPs. The electrodes were prepared with the variable growth time $t_2$ equal to (a) 1 min, (b) 5 min, (c) 10 min, (d) 20 min, and (e) 30 min. The SEM images were taken with horizontal field widths of 100 μm (top row), 25 μm (middle row), and 10 μm (bottom row). (f) Mid-infrared reflectance spectra of the same set of electrodes exhibiting a clear blue shift of the localized surface plasmon resonance wavelength with the decreasing average size of the MPs. The inset shows the zoom-in view of the same data.

To get a deeper insight into the nature of the plasmonic resonance modes within AgA MPs, it is necessary to obtain their dielectric function and resort to numerical simulations. Unfortunately, there are currently no reports of the optical properties of AgA in the literature. To fill this gap and to get a valid input for our calculations, we measured the dielectric function of a bulk AgA sample using spectroscopic ellipsometry (Figure 3a). Although the MIR wavelengths were beyond the accessible spectral range of our ellipsometer, our measured data still cover a wide domain from ultraviolet (UV) to near-infrared (NIR) frequencies. This prevents us to further discuss the response of AgA *MPs* shown above but provides a solid starting point for the analysis of AgA *NPs* that will be discussed in the following paragraphs.

The NIR optical properties of bulk AgA are dominated by high optical losses, which manifest themselves in the large value of the dielectric function's imaginary part. We put this result into somewhat better perspective by plotting the standard reference data for silver and gold[56] in Figure 3a as well. Note that the increasingly lossy behavior at long wavelengths is well in line with the observed broad plasmonic resonances supported by the MPs in the MIR range (cf. Figure 2f). At VIS and UV wavelengths, on the contrary, the losses in AgA should be on par or even better than those in the prototypical plasmonic materials like silver or gold. Also, the real part of the AgA dielectric function drops down to more negative values, indicating the potential of this material for plasmonic applications within this spectral range, which we are going to confirm below.

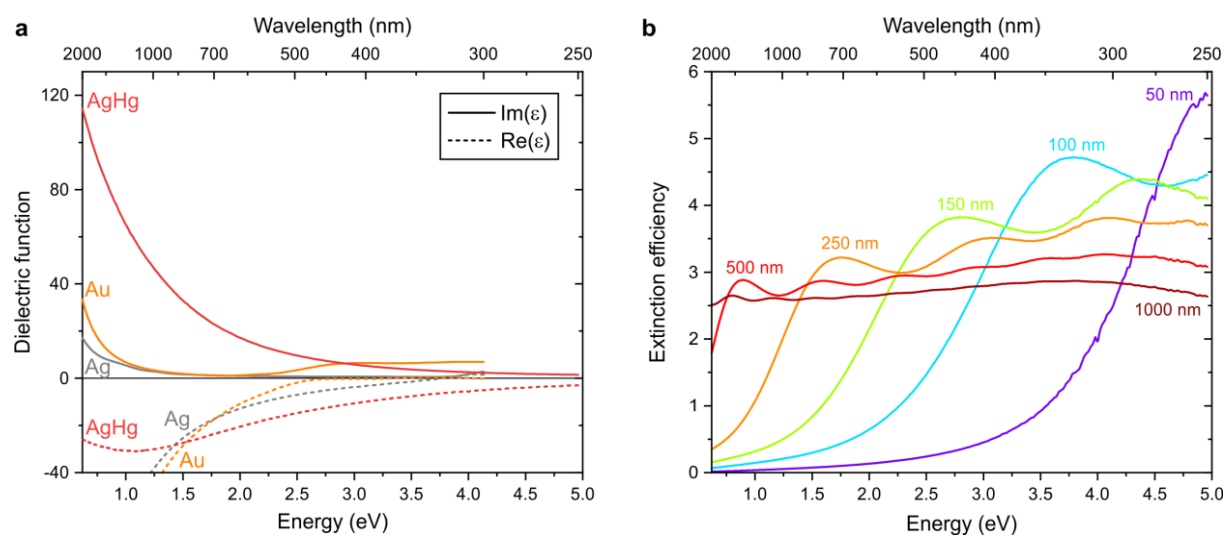

**Figure 3:** (a) Dielectric function of bulk AgA (red lines), as measured using spectroscopic ellipsometry. Literature data for silver (gray) and gold (orange) are also shown for reference.[56] (b) Theoretical calculation of the extinction efficiency spectra of spherical AgA NPs of various diameters (as labeled) with the material properties defined by the dielectric function obtained from the measurements shown in (a).

To study the general plasmonic properties of AgA NPs without any particular focus on their shape or environment, we use the classic Mie theory,[41,57] although FDTD simulations provide qualitatively the same results (Supporting Information, Figure S3). The single most informative attribute of the plasmonic resonance of any NP is probably the extinction cross-section or its normalized counterpart — the extinction efficiency ($Q_{ext}$).[58] Therefore, we modeled a set of spherical AgA NPs in vacuum, with the measured AgA dielectric function, and with the diameters varying between 50 nm and 1000 nm. The resulting $Q_{ext}$ spectra, presented in Figure 3b, reveal the characteristic red-shifting dipolar resonance, as well as the higher-order multipolar resonances at shorter wavelengths. Due to significant losses in the NIR range, plasmonic resonances of the large MPs are very broad, following the trend observed in the measured MIR spectra shown above. For the smaller NPs, however, the resonances in the VIS and UV ranges are of higher quality and strength. Overall, this straightforward theoretical prediction suggests that the small AgA NPs should exhibit a set of multimodal plasmonic resonances, which are worth further investigation. To this end, we carried out a series of EELS experiments on a single-particle level.

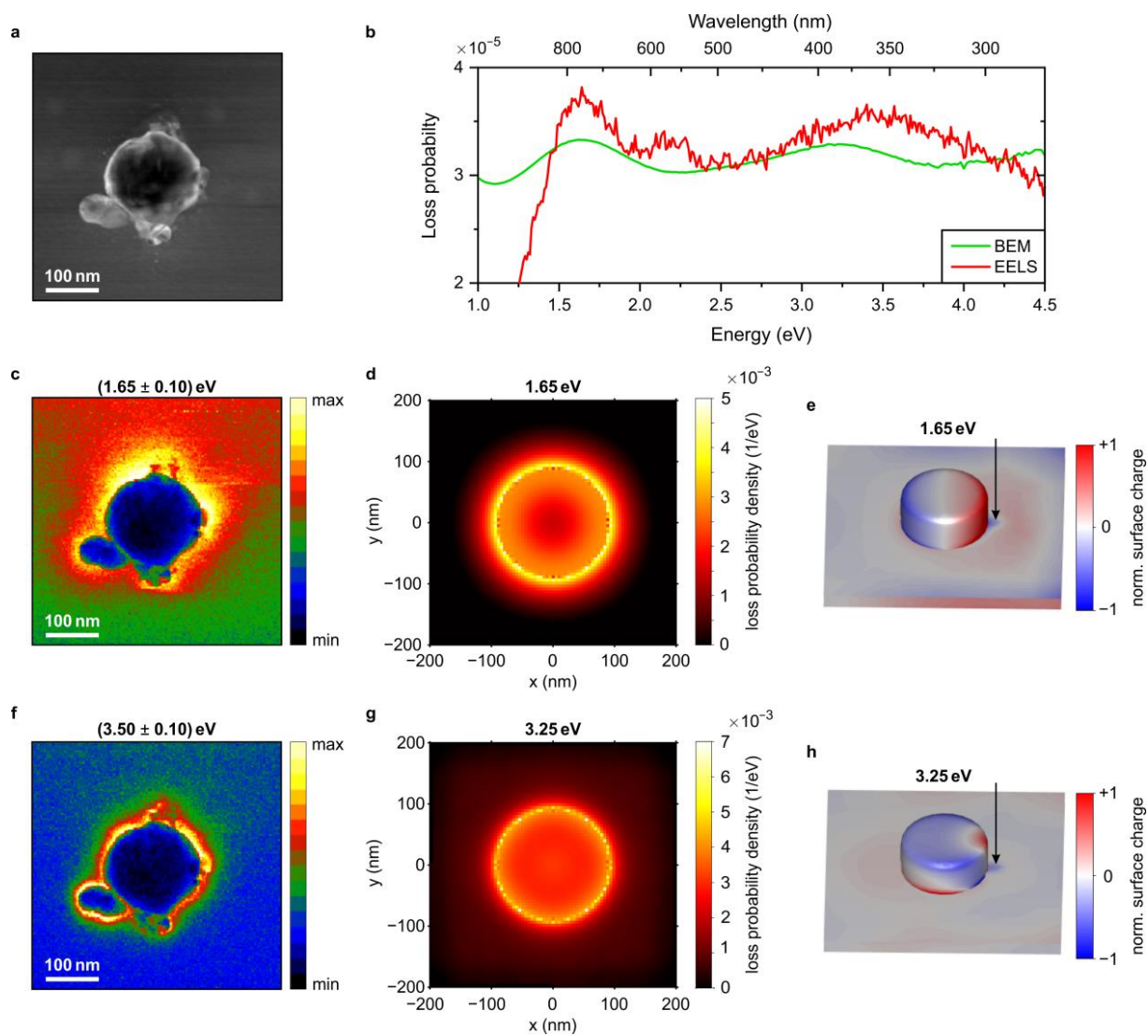

**Figure 4:** (a) SEM image of a AgA NP selected for EELS analysis. (b) Electron energy loss probability spectra of the same NP — both measured using EELS (red) and calculated by BEM (green). (c) Measured EELS map of the NP, energy-filtered at 1.65 eV. (d) Calculated electron energy loss probability density and (e) normalized surface charge distribution at the same energy. The direction of the incident electron beam is marked by the black arrow. (f)–(h) Same as (c)–(e) but at the spectral positions of the high-energy mode.

We chose EELS for its excellent spatial resolution, wide accessible spectral range from UV to NIR, and ability to disentangle different plasmonic modes in a straightforward manner.[59] The single disc-shaped AgA NP selected for investigation is shown in Figure 4a, while analysis of NPs with other shapes and sizes can be found in the Supporting Information, Figure S4. It is worth noting that AgA NPs are very often surrounded by a small number of significantly smaller NPs with a very high silver content (Supporting Information, Figure S5). We ascribe their presence to be an artifact of the NP transfer process to the TEM membrane, as they are not present on the as-grown samples on the top of the ITO layer. The EEL spectrum of the selected NP is shown in Figure 4b, exhibiting two resonance peaks at 1.65 and 3.50 eV. The corresponding EEL maps, energy-filtered at the two resonance energies, are presented in Figure 4c,f. To elucidate the nature of the observed plasmonic resonances, we employed BEM numerical simulations and modeled the NP as a disc with a diameter of 175 nm and thickness of 80 nm on top of a 30 nm thick silicon nitride membrane. The resulting calculated loss probability spectrum is plotted over the measured one in Figure 4b, with a reasonably good match given the considerable simplifications with respect to particle shape and uncertainty in the material properties. The calculated EEL maps shown in Figure 4d,g also reproduce the main features of the measured ones very well, i.e., they reveal the tight spatial confinement of the high-energy resonance. The clearest description of the plasmonic modes can be usually obtained from the surface charge distribution, as demonstrated in Figure 4e, where the dipolar nature of the low-energy plasmonic mode can be clearly observed. The calculated surface charge distribution of the high-energy mode (Figure 4h), however, unveils a distorted and ambiguous resonant mode. The simulations of a bare AgA nanodisc without the suspected source of the distortion — the silicon nitride membrane — reveal the clear quadrupolar nature of this mode (Supporting Information, Figure S6), as expected due to its energy being approximately the double of the dipolar mode. We confirmed the similar plasmonic behavior on AgA NPs of other sizes and shapes. For AgA nanorods, for example, we observed the classic couple of a transversal and a longitudinal resonance, partially overlapping with a quadrupolar mode (Supporting Information, Figure S7).

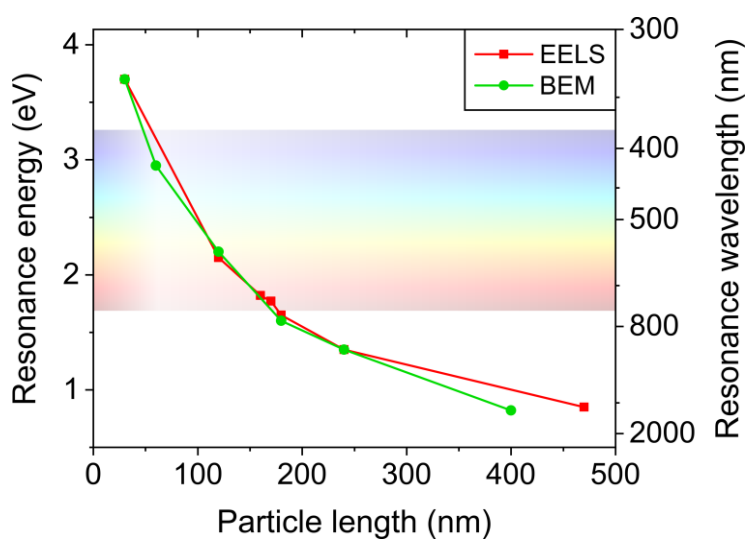

**Figure 5:** Measured (EELS) and calculated (BEM) plasmonic resonance energies corresponding to the dipole modes of a set of AgA nanoparticles spanning the region from UV across VIS (marked by the rainbow area in the background) to NIR.

An important aspect, with regards to potential applications, is the spectral range that can be covered by the plasmonic resonances of AgA NPs. We therefore analyzed the measured spectral positions of the longitudinal plasmonic resonances of a series of AgA NPs with their nominal sizes from 30 to 470 nm (Supporting Information, Figure S4). The extracted resonance energies are shown in Figure 5 as a function of the particle size, complemented by the corresponding results of the BEM numerical simulations. The observed resonances span the extremely wide spectral region from ultraviolet, across visible, up to near-infrared wavelengths. Together with the MIR resonances studied above (Figure 2), these results render the AgA NPs and MPs a promising plasmonic material with extraordinary tunability, which will be especially useful for applications in the realm of photochemistry and spectroelectrochemistry.

**CONCLUSIONS**

In conclusion, we have presented how electrodeposition from a mixture of silver and mercury precursors leads to the growth of high-quality solid AgA NPs and MPs. We have verified, using both optical and EEL spectroscopies, that these particles support broad plasmonic resonances, which can be tuned by control of their nominal size. These resonances cover an extremely wide spectral window from the MIR for the largest particles down to the UV for the smallest ones. Our results are supported by numerical simulations, using the measured AgA dielectric function and real particle sizes as the model parameters. With respect to the potential applications, previous works reported that AgA exhibits extraordinary electrochemical properties, its cathodic potential window is exceptionally wide, and biomolecules like DNA or proteins exhibit enhanced affinity to its surface. The unique combination of these electrochemical properties with the wide spectrally tunable plasmonic functionality reported here makes AgA particles an interesting platform for future spectroelectrochemical or photocatalytic applications.

## ASSOCIATED CONTENT

**Supporting Information:** Size distribution statistics; mercury weight percentage in the AgA MPs; FDTD numerical simulations; TEM images and EEL maps of additional AgA NPs; EDX maps of AgA NPs with Ag satellites; numerical simulations of a AgA nanodisc in vacuum and EELS and BEM analysis of a AgA nanorod.

## AUTHOR INFORMATION

**Corresponding Author**

*E-mail: filip.ligmajer@ceitec.vutbr.cz (F.L.).

**ORCID**

Filip Ligmajer: 0000-0003-0346-4110

Michal Horák: 0000-0001-6503-8294

Miroslav Fojta: 0000-0003-1854-6506

Aleš Daňhel: 0000-0003-4917-3263

**Notes**

The authors declare no competing financial interest. The data created in this research are publicly available on the Figshare web site at https://doi.org/ 10.6084/m9.figshare.8020295.

## ACKNOWLEDGMENTS

This work has been supported by the Grant Agency of the Czech Republic (projects 17-23634Y and 17-33767L) and by the SYMBIT project reg. no. CZ.02.1.01/0.0/0.0/15_003/0000477 financed from the ERDF. A part of this work was carried out with the support of the MEYS CR under the National Sustainability Programme II (project CEITEC 2020, LQ1601), CEITEC Nano Research Infrastructure (ID LM2015041, MEYS CR, 2016–2019), and Brno University of Technology (project CEITEC VUT-J-19-5945).

**TOC IMAGE**

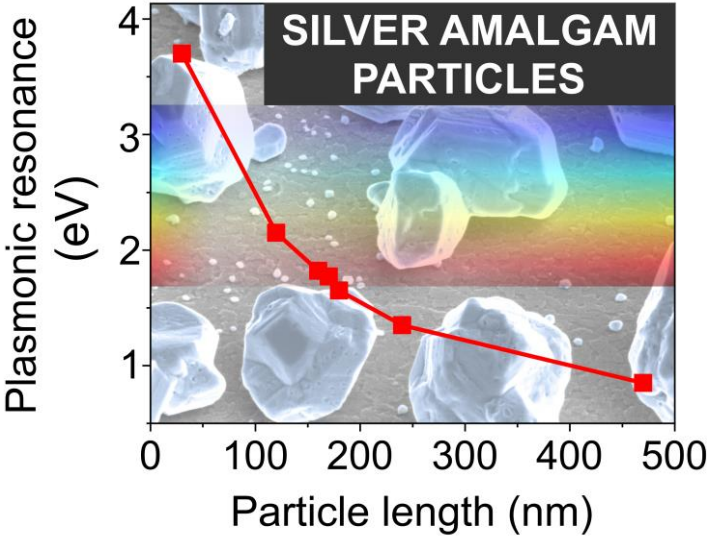

Supporting Information

for

# Silver Amalgam Nanoparticles and Microparticles: A Novel Plasmonic Platform for Spectroelectrochemistry

Filip Ligmajer,*,[†,‡,§] Michal Horák,[†,§] Tomáš Šikola,[†,§] Miroslav Fojta,[‡,∥] Aleš Daňhel[‡]

[†] *Central European Institute of Technology, Brno University of Technology, Purkyňova 123, Brno, 612 00, Czech Republic*

[‡] *Institute of Biophysics of the Czech Academy of Sciences, Královopolská 135, Brno, 612 65, Czech Republic*

[§] *Institute of Physical Engineering, Faculty of Mechanical Engineering, Brno University of Technology, Technická 2, Brno, 616 69, Czech Republic*

[∥] *Central European Institute of Technology, Masaryk University, Kamenice 753/5, Brno, 625 00, Czech Republic*

**Corresponding Author**

*E-mail: filip.ligmajer@ceitec.vutbr.cz (F.L.).

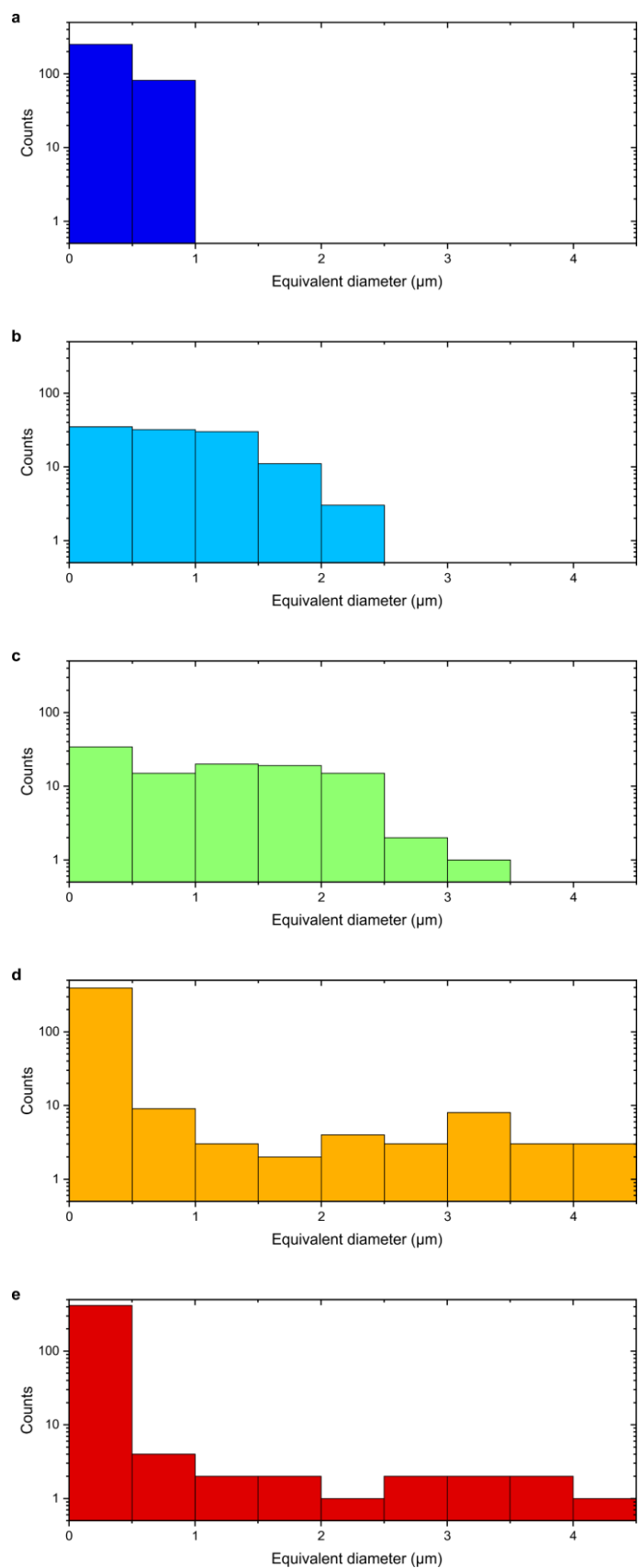

**Figure S1:** Histogram of AgA MPs' size distribution on the electrodes prepared with the variable growth time $t_2$ equal to (a) 1 min, (b) 5 min, (c) 10 min, (d) 20 min, (e) 30 min. The particles' equivalent diameters were obtained using an image intensity thresholding algorithm run on the SEM images of the electrodes.

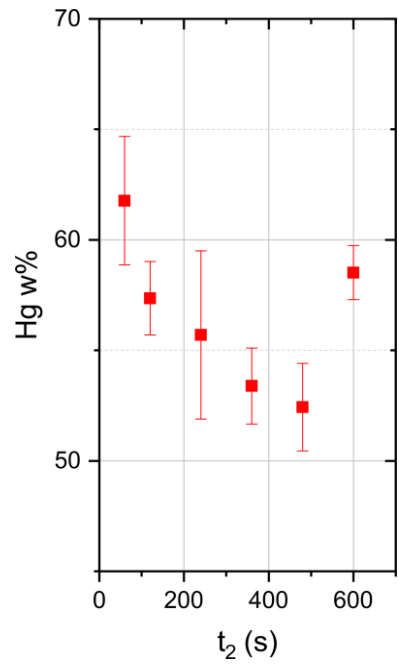

**Figure S2:** Mercury weight percentage (Hg w%) in the AgA MPs as a function of their growth time ($t_2$). The data were extracted from point EDS spectra taken from at least 10 MPs on each sample.

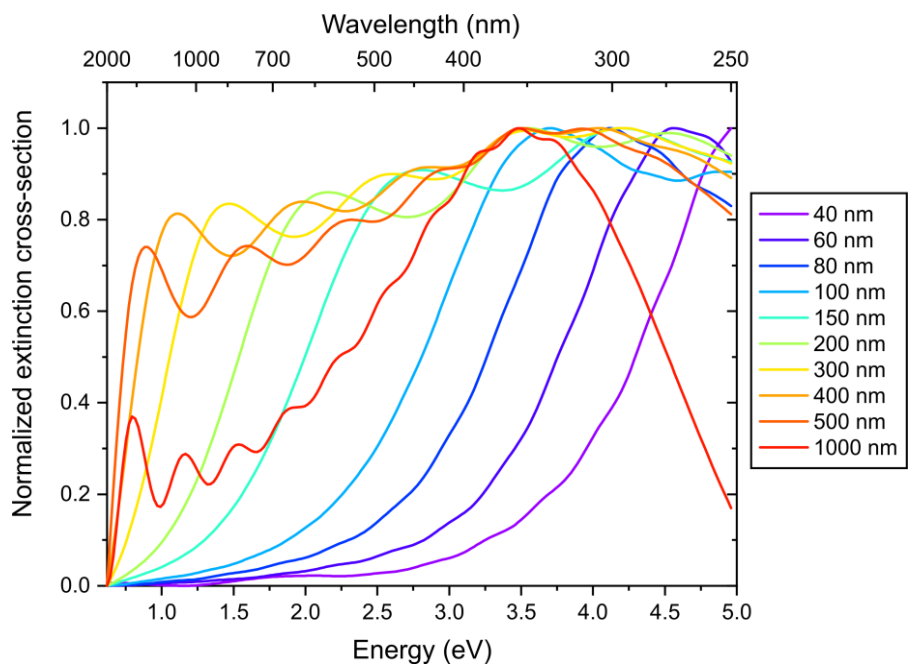

**Figure S3:** FDTD numerical simulations of spherical AgA NPs and NPs of various diameters (as labeled) placed in vacuum, with their material properties defined using the dielectric function obtained by spectroscopic ellipsometry.

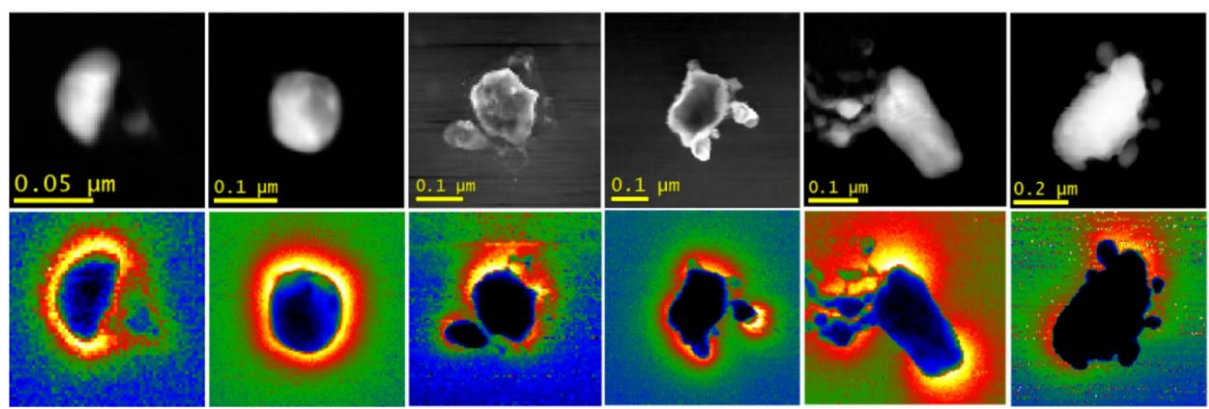

**Figure S4:** High-angle annular dark-field (HAADF) scanning transmission electron micrographs of AgA NPs with the nominal sizes ranging from 30 nm to 470 nm (top) and corresponding electron energy loss maps at the energy corresponding to their dipolar mode (bottom).

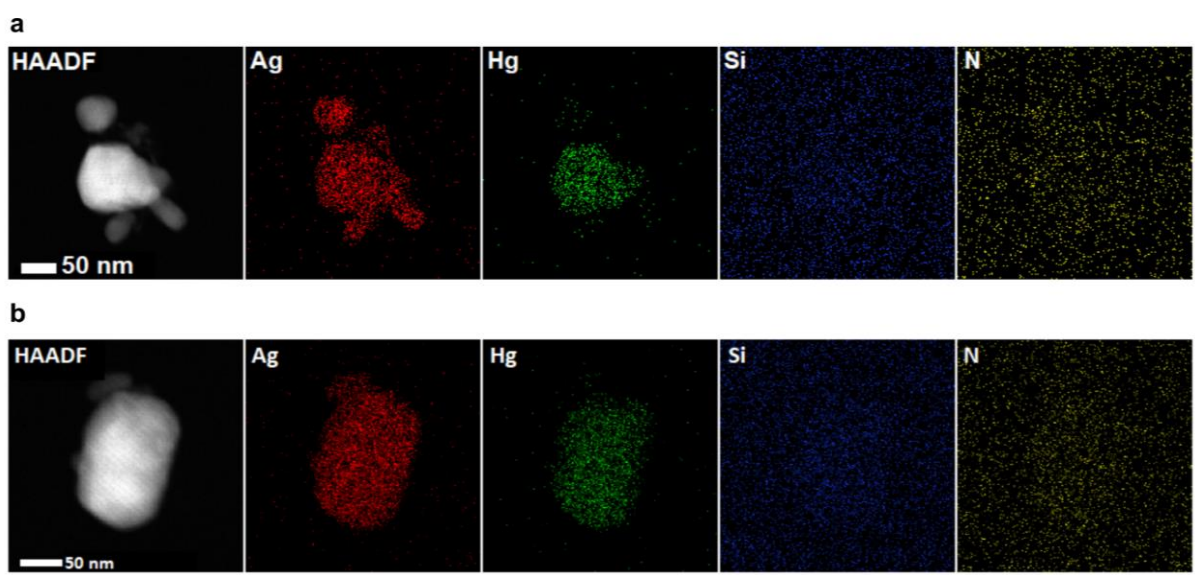

**Figure S5:** (a,b) HAADF scanning transmission electron micrograph of two representative AgA NPs with the characteristic satellite silver nanoparticles located in their vicinity. The corresponding elemental maps from energy dispersive X-ray spectroscopy analysis verify the negligible mercury content in the satellites. The composition of the satellites was estimated to average silver weight percentage of (62 ± 4) %.

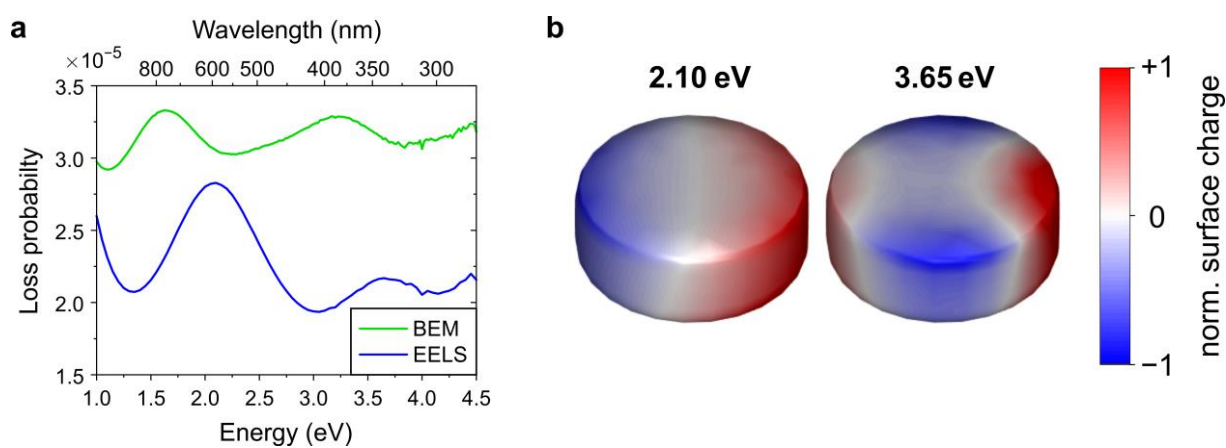

**Figure S6:** (a) Calculated EEL probability spectra for an AgA nanodisc (175 nm in diameter, 80 nm thick) on top of a silicon nitride membrane (green) and placed in vacuum (blue). (b) Calculated normalized surface charge distribution for the same disc at the energy of the dipolar mode (2.10 eV) and the quadrupolar mode (3.65 eV).

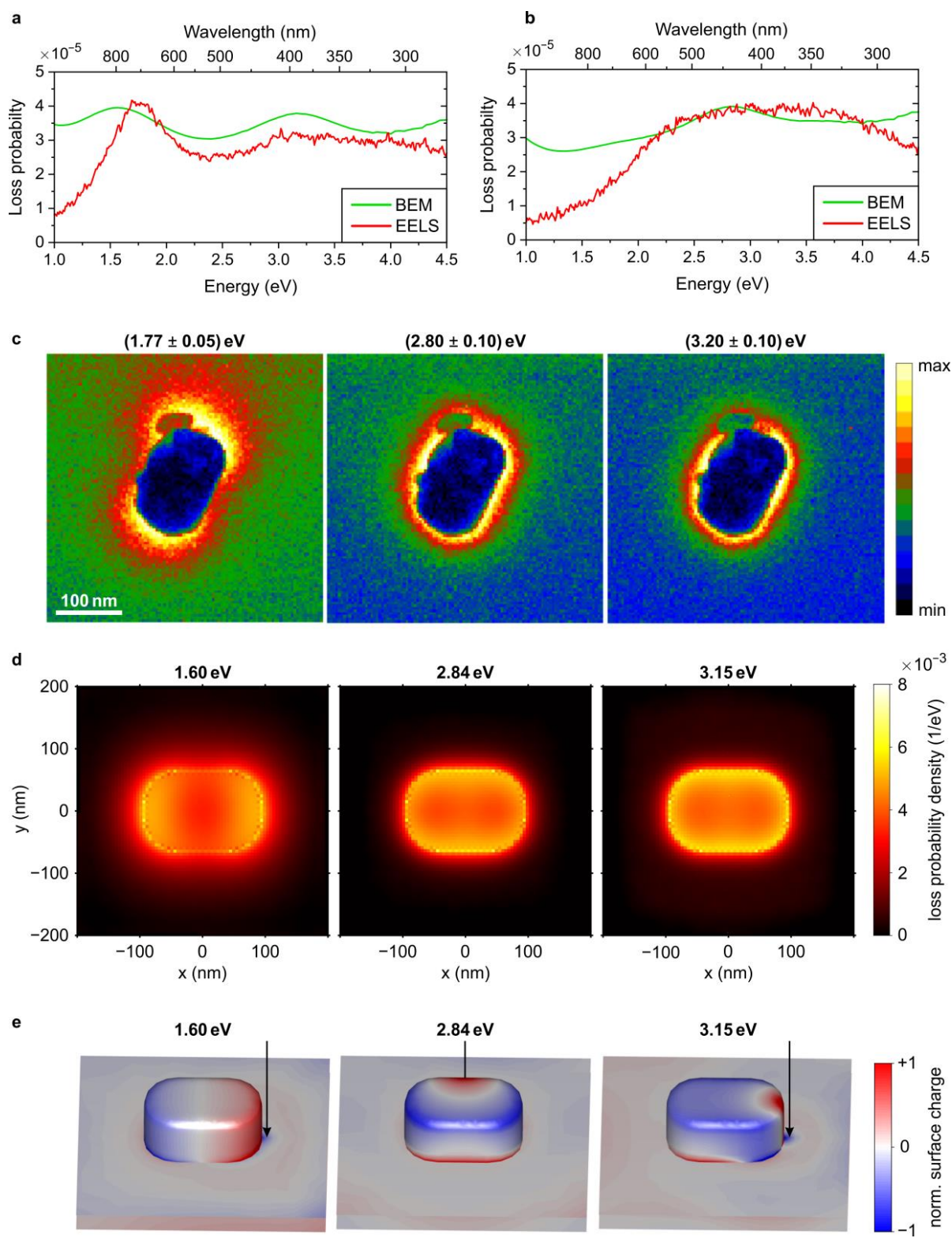

**Figure S7:** (a,b) EEL probability spectra of the AgA NPs shown in Figure S4b — both measured using EELS (red) and calculated by BEM (green) for the excitation near its short edge (a) and long edge (b). (c) Measured EELS map of the same NP, energy filtered around the three modes of its plasmonic resonance. (d) Calculated EEL probability density and (e) normalized surface charge distribution at the energies of the calculated resonance peaks. The direction of the incident electron beam is marked by the black arrows.

29